# Confounding Equivalence in Causal Inference

**Judea Pearl**
Cognitive Systems Laboratory
Computer Science Department
University of California, Los Angeles, CA 90024 USA
*judea@cs.ucla.edu*

**Azaria Paz**
Department of Computer Science
Technion IIT, Haifa 3200
*paz@cs.Technion.AC.IL*

## Abstract

The paper provides a simple test for deciding, from a given causal diagram, whether two sets of variables have the same bias-reducing potential under adjustment. The test requires that one of the following two conditions holds: either (1) both sets are admissible (i.e., satisfy the back-door criterion) or (2) the Markov boundaries surrounding the manipulated variable(s) are identical in both sets. Applications to covariate selection and model testing are discussed.

## 1 INTRODUCTION

The common method of estimating causal effects in observational studies is to *adjust* for a set of variables (or "covariates") judged to be "confounders," that is, variables capable of producing spurious associations between treatment and outcome, not attributable to their causal dependence. While adjustment tends to reduce the bias produced by such spurious associations, the bias-reducing potential of any set of covariates depends crucially on the causal relationships among all variables affecting treatment or outcome, hidden as well as visible. Such relationships can effectively be represented in the form of directed acyclic graphs (DAG's) (Pearl, 1995; Lauritzen, 2001; Spirtes et al., 2000; Glymour and Greenland, 2008; Dawid, 2002).

Most studies of covariate selection have aimed to define and identify "admissible" sets of covariates, also called "sufficient sets," namely, a set of covariates that, if adjusted for, would yield asymptotically unbiased estimates of the causal effect of interest (Stone, 1993; Greenland et al., 1999; Pearl, 2000). A graphical criterion for selecting an admissible set is given by the "back-door" test (Pearl, 1993, 2000) which was shown to entail zero bias, or "no confoundedness," assuming correctness of the causal assumptions encoded in the DAG. Related notions are "exchangeability" (Greenland and Robins, 1986), "exogeneity" (Engle et al., 1983), and "strong ignorability" (Rosenbaum and Rubin, 1983).

This paper addresses a different question: Given two sets of variables in a DAG, decide if the two are equally valuable for adjustment, namely, whether adjustment for one set is guaranteed to yield the same asymptotic bias as adjustment for the other.

The reasons for posing this question are several. First, an investigator may wish to assess, prior to taking any measurement, whether two candidate sets of covariates, differing substantially in dimensionality, measurement error, cost, or sample variability are equally valuable in their bias-reduction potential. Second, assuming that the structure of the underlying DAG is only partially known, one may wish to assess, using $c$-equivalence tests, whether a given structure is compatible with the data at hand; structures that predict equality of post-adjustment associations must be rejected if, after adjustment, such equality is not found in the data.

In Section 2 we define $c$-equivalence and review the auxiliary notions of admissibility, $d$-separation, and the back-door criterion. Section 3 derives statistical and graphical conditions for $c$-equivalence, the former being sufficient while the latter necessary and sufficient. Section 4 presents a simple algorithm for testing $c$-equivalence, while Section 5 gives a statistical interpretation to the graphical test of Section 3. Finally, Section 6 discusses potential applications of $c$-equivalence for effect estimation, model testing, and model selection.

## 2 PRELIMINARIES: c-EQUIVALENCE AND ADMISSIBILITY

Let $X, Y$, and $Z$ be three disjoint subsets of discrete variables, and $P(x, y, z)$ their joint distribution. We are concerned with expressions of the type

$$A(x, y, Z) = \sum_z P(y|x, z) P(z) \qquad (1)$$

Such expressions, which we name "adjustment estimands," are often used to approximate the causal effect of $X$ on $Y$, where the set $Z$ is chosen to include variables judged to be "confounders." By *adjusting* for these variables, one hopes to create conditions that eliminate spurious dependence and thus obtain an unbiased estimate of the causal effect of $X$ and $Y$, written $P(y|do(x))$ (see Pearl (2000) for formal definition and methods of estimation).

**Definition 1.** *(c-equivalence)*
*Define two sets, $T$ and $Z$ as c-equivalent (relative to $X$ and $Y$), written $T \approx Z$, if the following equality holds for every $x$ and $y$:*

$$\sum_t P(y|x, t) P(t) = \sum_z P(y|x, z) P(z) \qquad \forall\ x, y \quad (2)$$

*or*
$$A(x, y, T) = A(x, y, Z) \qquad \forall\ x, y$$

*This equality guarantees that, if adjusted for, sets $T$ and $Z$ would produce the same asymptotic bias relative to the target quantity.*

**Definition 2.** *(Admissibility)*
*Let $P(y|do(x))$ stand for the "causal-effect" of $X$ on $Y$, i.e., the distribution of $Y$ after setting variable $X$ to a constant $X = x$ by external intervention. A set $Z$ of covariates is said to be "admissible" (for adjustment) relative to the causal effect of $X$ on $Y$, if the following equality holds:*

$$\sum_z P(y|x, z) P(z) = P(y|do(x)) \qquad (3)$$

Equivalently, one can define admissibility using the equalities:
$$P(y|do(x)) = P(Y_x = y) \qquad (4)$$

where $Y_x$ is the counterfactual or "potential outcome" variable (Neyman, 1923; Rubin, 1974). The equivalence of the two definitions is shown in (Pearl, 2000).

**Definition 3.** *(d-separation)*
*A set $S$ of nodes in a graph $G$ is said to block a path $p$ if either (i) $p$ contains at least one arrow-emitting node that is in $S$, or (ii) $p$ contains at least one collision node that is outside $S$ and has no descendant in $S$. If $S$ blocks all paths from $X$ to $Y$, it is said to "d-separate $X$ and $Y$," written $(X \perp\!\!\!\perp Y | S)_G$ and then, $X$ and $Y$ are independent given $S$, written $X \perp\!\!\!\perp Y | S$, in every probability distribution that can be generated by a process structured along $G$ (Pearl, 1988).*

**Lemma 1.** *(The back-door criterion: G-admissibility) Let $G$ be a directed acyclic graph (DAG) that encodes the causal relationships between variables in a problem, observables as well as unobservable. A sufficient condition for a subset $S$ of covariates to be admissible is that it satisfies the following two conditions (Pearl, 1993):*

1. *No element of $S$ is a descendant of $X$*
2. *The elements of $S$ "block" all "back-door" paths from $X$ to $Y$, namely all paths that end with an arrow pointing to $X$.*

*A set $S$ satisfying the back-door criterion will be called G-admissible.*

For proof and intuition behind the back-door test, especially a relaxation of the requirement of no descendants, see (Pearl, 2009a, p. 339).

Clearly, if two subsets $Z$ and $T$ are G-admissible they are also admissible and they must be c-equivalent, for their adjustment estimands coincide with the causal effect $P(y|do(x))$. Therefore, a trivial graphical condition for c-equivalence is for $Z$ and $T$ to satisfy the back-door criterion of Lemma 1. This condition, as we shall see in the next section, is rather weak; c-equivalence extends beyond admissible sets.

## 3 CONDITIONS FOR c-EQUIVALENCE

**Theorem 1.** *A sufficient condition for the c-equivalence of $T$ and $Z$ is that $Z$ satisfies:*

$$\begin{array}{ll}(X \perp\!\!\!\perp Z | T) & (i) \\ (Y \perp\!\!\!\perp T | X, Z) & (ii)\end{array}$$

**Proof:**
Conditioning on $Z$, $(ii)$ permits us to rewrite the left-hand side of (2) as

$$\begin{array}{ll}A(x, y, T) & = \sum_t P(t) \sum_z P(y|z, x, t) P(z|t, x) \\ & = \sum_t P(t) \sum_z P(y|z, x) P(z|t, x)\end{array}$$

and $(i)$ further yields $P(z|t, x) = P(z|t)$, from which (2) follows:

$$\begin{array}{ll}A(x, y, T) & = \sum_t \sum_z P(y|z, x) P(z, t) \\ & = \sum_z P(y|z, x) P(z) \\ & = A(x, y, Z)\end{array}$$

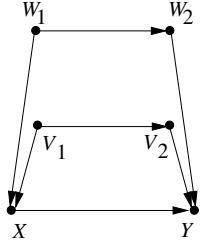

Figure 1: The sets $T = \{V_1, W_1\}$ and $Z = \{V_2, W_2\}$ satisfy the conditions of Theorem 1. The sets $T = \{V_1, W_2\}$ and $Z = \{V_2, W_2\}$ block all back-door paths between $X$ and $Y$, hence they are admissible and $c$-equivalent. Still they do not satisfy the conditions of Theorem 1.

**Corollary 1.** *A sufficient condition for the $c$-equivalence of $T$ and $Z$ is that either one of the following two conditions holds:*

$$\begin{aligned} C^* : \quad & X \perp\!\!\!\perp Z | T \quad \text{and} \quad Y \perp\!\!\!\perp T | Z, X \\ C^{**} : \quad & X \perp\!\!\!\perp T | Z \quad \text{and} \quad Y \perp\!\!\!\perp Z | T, X \end{aligned} \quad (5)$$

**Proof:**
$C^*$ permits us to derive the right hand side of Eq. (2) from the left hand side, while $C^{**}$ permits us to go the other way around.

The conditions offered by Theorem 1 and Corollary 1 do not characterize *all* equivalent pairs, $T$ and $Z$. For example, consider the graph in Figure 1, in which each of $T = \{V_1, W_2\}$ and $Z = \{V_2, W_1\}$ is $G$-admissible they must therefore be $c$-equivalent. Yet neither $C^*$ nor $C^{**}$ holds in this case.

On the other hand, condition $C^*$ can detect the $c$-equivalence of some non-admissible sets, such as $T = \{W_1\}$ and $Z = \{W_1, W_2\}$. These two sets are non-admissible for they fail to block the back-door path $X \leftarrow V_1 \rightarrow V_2 \rightarrow Y$, yet they are $c$-equivalent according to Theorem 1; (i) is satisfied by $d$-separation, while (ii) is satisfied by subsumption ($T \subseteq Z$).

It is interesting to note however that $Z = \{W_1, W_2\}$, while $c$-equivalent to $\{W_1\}$, is not $c$-equivalent to $T = \{W_2\}$, though the two sets block the same path in the graph.[1] Indeed, this pair does not meet the test of Theorem 1; choosing $T = \{W_2\}$ and $Z = \{W_1, W_2\}$ violates condition (i) since $X$ is not $d$-separated from $W_1$, while choosing $Z = \{W_2\}$ and $T = \{W_1, W_2\}$ violates condition (ii) by unblocking the path $W_1 \rightarrow X \leftarrow V_1 \rightarrow V_2 \rightarrow Y$. Likewise, the sets $T = \{W_1\}$ and $Z = \{W_2\}$ block the same path and, yet, are not $c$-equivalent; they fail indeed to satisfy condition (ii) of Theorem 1.

---

[1] The reason is that the strength of the association between $X$ and $Y$, conditioned on $W_2$, depends on whether we also condition on $W_1$. Else, $P(y|x, w_2)$ would be equal to $P(y|x, w_1, w_2)$ which would render $Y$ and $W_1$ independent given $X$ and $W_2$. But this is true only if the path $(X, V_1, V_2, Y)$ is blocked.

We are now ready to broaden the scope of Theorem 1 and derive a condition (Theorem 2 below) that detects *all* $c$-equivalent subsets in a graph.

**Definition 4.** *(Markov boundary)*
*For any set of variables $S$ in $G$; let $S_m$ be the minimal subset of $S$ that satisfies the condition*

$$(X \perp\!\!\!\perp S | S_m)_G \quad (6)$$

*In other words, measurement of $S_m$ renders $X$ independent of all other members of $S$, and no proper subset of $S_m$ has this property. This minimal subset is called the Markov Boundary of $X$ relative to $S$, or "Markov boundary" for short.*

The Markov boundary as defined above is known to be unique, which follows from the intersection property of $d$-separation (Pearl, 1988, p. 97). See Appendix A.

**Lemma 2.** *Every set of variables, $S$, is $c$-equivalent to its Markov boundary $S_m$.*

**Proof.**
Choosing $Z = S$ and $T = S_m$ satisfies the two conditions of Theorem 1; (i) is satisfied by the definition of $S_m$, while (ii) is satisfied by subsumption ($T \subseteq Z$).

**Theorem 2.** *Let $Z$ and $T$ be two sets of variables containing no descendant[2] of $X$. A necessary and sufficient conditions for $Z$ and $T$ to be $c$-equivalent is that at least one of the following conditions holds:*
1. $Z_m = T_m$
*or*
2. $Z$ *and* $T$ *are $G$-admissible*

**Proof**

1. Proof of sufficiency:

    Condition 2 is sufficient since $G$-admissibility implies admissibility and renders the two adjustment estimands in (2) equal to the causal effect. Condition 1 is sufficient by reason of Lemma 2, which

---

[2] The sufficiency part of Theorem 2 is valid without excluding descendants of $X$. The necessary part can allow descendants of $X$ if we interpret $Z_m$ and $T_m$ to represent "reduced Markov boundaries," namely Markov boundaries exclusive of all nodes that are $d$-separated from $Y$ given $X$. For example, two children of $X$ that are not on pathways to $Y$ would then be properly identified as "$c$-equivalent," even though they do not satisfy the conditions of Theorem 2.

yields:
$$Z \approx Z_m \approx T_m \approx T$$

2. Proof of necessity:
   We need to show that if conditions (1) and (2) are violated then there is at least one parameterization of $G$ (that is, an assignment of conditional probabilities to the parent-child families in $G$) that violates Eq. (2). If exactly one of $(Z, T)$ is $G$-admissible then $Z$ and $T$ are surely not $c$-equivalent, for their adjustment estimands would differ for some parameterization of the graph. Assume that both $Z$ and $T$ are not $G$-admissible or, equivalently, that none of $Z_m$ or $T_m$ is $G$-admissible. Then there is a back-door path $p$ from $X$ to $Y$ that is not blocked by either $Z_m$ or $T_m$. If, in addition, condition (1) is violated (i.e., $Z_m$ differs from $T_m$) then $T_m$ and $Z_m$ cannot both be disconnected from $X$, (for then $Z_m = T_m = \emptyset$, satisfying condition (1)), there must be a path $p_1$ from either $Z_m$ to $X$ that is not blocked by $T_m$ or a path $p_2$ from $T_m$ to $X$ that is not blocked by $Z_m$. Assuming the former case, there must be an unblocked path $p_1$ from $Z_m$ to $X$ followed by a back door path $p$ from $X$ to $Y$. The existence of this path implies that conditional on $t$, $Z$ acts as an instrumental variable with respect to the pair $(X, Y)$. For such a structure, the following parameterization would violate Eq. (2). First we weaken the links from $T_m$ to $X$ to make the left hand side of (2) equal to $P(y|x)$, or $A(x, y, T_m) = A(x, y, 0)$. Next, we construct a linear structural equation model in which $Z_m$ is a strong predictor of $X$ and $X$ and $Y$ are confounded. Wooldridge (2009) has shown (see also (Pearl, 2009b)). that adjustment on $Z_m$ under such conditions results in a higher bias relative to the unadjusted estimand, or $A(x, y, Z_m) \neq A(x, y, 0)$. This violates the equality in Eq. (2) and prove the necessary part of Theorem 2.

## 4 STEP-WISE TEST FOR $c$-EQUIVALENCE

Figure 2 illustrates the power of Theorem 2. In this model, no subset of $\{W_1, W_2, W_3\}$ is $G$-admissible (because of the back-door path through $V_1$ and $V_2$) and, therefore, equality of Markov boundaries is necessary and sufficient for $c$-equivalence among any two such subsets. Accordingly, we can conclude that $T = \{W_1, W_2\}$ is $c$-equivalent to $Z = \{W_1, W_3\}$, since $T_m = W_1$ and $Z_m = W_1$. Note that $W_1$ and $W_2$, though they result (upon conditioning) in the same set of unblocked paths between $X$ and $Y$, are not $c$-equivalent since $T_m = W_1 \neq Z_m = W_2$. Indeed, each of $W_1$ and $W_2$ is an instrumental variable relative to $\{X, Y\}$, with potentially different strengths, hence potentially different adjustment estimands. Sets $W_4$ and $W_3$ however are $c$-equivalent, because the Markov boundary of each is the null set, $\{\emptyset\}$.

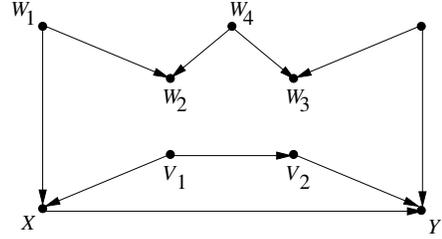

Figure 2: $W_3$ and $W_4$ are non-admissible yet $c$-equivalent; both having $\emptyset$ as a Markov boundary. However, $W_2$ and $W_3$ are not $c$-equivalent with Markov boundaries $W_2$ and $\emptyset$, respectively.

We note that testing for $c$-equivalence can be accomplished in polynomial time. The Markov boundary of an arbitrary set $S$ can be identified by iteratively removing from $S$, in any order, any node that is $d$-separated from $X$ given all remaining members of $S$ (see Appendix A). $G$-admissibility, likewise, can be tested in polynomial time (Tian et al., 1998).

Theorem 2 also leads to a step-wise process of testing $c$-equivalence,

$$T \approx T_1 \approx T_2 \approx \ldots \approx Z$$

where each intermediate set is obtained from its predecessor by an addition or deletion of one variable only. This can be seen by organizing the chain into three sections.

$$T \approx \ldots \approx T_m \approx \ldots \approx Z_m \approx \ldots \approx Z$$

The transition from $T$ to $T_m$ entails the deletion from $T$ of all nodes that are not in $T_m$; one at a time, in any order. Similarly, the transition from $Z_m$ to $Z$ builds up the full set $Z$ from its Markov boundary $Z_m$; again, in any order. Finally, the middle section, from $T_m$ to $Z_m$, amounts to traversing a chain of $G$-admissible sets, using both deletion and addition of nodes, one at a time. A Theorem due to (Tian et al., 1998) ensures that such a step-wise transition is always possible between any two $G$-admissible sets. In case $T$ or $Z$ are non-admissible, the middle section must degenerate into an equality $T_m = Z_m$, or else, $c$-equivalence does not hold.

Figure 2 can be used to illustrate this stepwise transition from $T = \{W_1, W_2, V_1\}$ to $Z = \{V_2, W_3\}$. Start-

ing with $T$, we obtain:

$$T = \{W_1, W_2, V_1\} \approx \{W_1, V_1\} = T_m \approx \{V_1\} \approx \{V_1, V_2\}$$
$$\approx \{V_2\} = Z_m \approx \{V_2, W_3\} = Z$$

If, however we were to attempt a stepwise transition between $T = \{W_1, W_2, V_1\}$ and $Z = \{W_3\}$, we would obtain:

$$T = \{W_1, W_2, V_1\} \approx \{W_1, V_1\} \approx \{V_1\}$$

and would be unable to proceed toward $Z_m = \{W_3\}$. The reason lies in the non-admissibility of $Z$ which necessitates the equality $T_m = Z_m$, contrary to Markov boundaries shown in the graph.

Note also that each step in the process $T \approx \ldots \approx T_m$ (as well as $Z_m \approx \ldots \approx Z$) is licensed by condition (i) of Theorem 1, while each step in the intermediate process $T_m \approx \ldots \approx Z_m$ is licensed by condition (ii). Both conditions are purely statistical and do not invoke the causal reading of "admissibility." This means that condition 2 of Theorem 2 may be replaced by the requirement that $Z$ and $T$ satisfy the back-door test in any diagram compatible[3] with $P(x, y, z, t)$; the direction of arrows in the diagram need not convey causal information. Further clarification of the statistical implications of the admissibility condition, is given in the next section.

## 5 FROM CAUSAL TO STATISTICAL CHARACTERIZATION

Theorem 2, while providing a necessary and sufficient condition for $c$-equivalence, raises an interesting theoretical question. Admissibility is a causal notion, (i.e., resting on causal assumptions about the direction of the arrows in the diagram) while $c$-equivalence is purely statistical. Why need one resort to causal assumptions to characterize a property that relies on no such assumption? Evidently, the notion of $G$-admissibility as it was used in the proof of Theorem 2 was merely a surrogate carrier of statistical information; its causal reading was irrelevant. The question then is whether Theorem 2 could be articulated using purely statistical conditions, avoiding admissibility altogether, as is done in Theorem 1.

We will show that the answer is positive; Theorem 2 can be rephrased using a statistical test for $c$-equivalence. It should be noted though, that the quest for statistical characterization is of merely theoretical

---

[3]A diagram $G$ is said to be compatible with a probability function $P$ if there is a parameterization of $G$ that generates $P$, i.e., if every $d$-separation in $G$ corresponds to a valid conditional independence in $P$.

---

interest; rarely is one in possession of prior information about conditional independencies, (as required by Theorem 1), that is not resting on causal knowledge (of the kind required by Theorem 2). The utility of statistical characterization surfaces when we wish to confirm or reject the structure of the diagram. We will see that the statistical reading of Theorem 2 has testable implication that, if failed to fit the data, may help one select among competing graph structures.

Our first step is to apply Theorem 1 to the special case where $T$ is a subset of $Z$.

**Theorem 3.** *(Set-subset equivalence)*
Let $T$ and $S$ be two disjoint sets. A sufficient condition for the $c$-equivalence of $T$ and $Z = T \cup S$ is that $S$ can be partitioned into two subsets, $S_1$ and $S_2$, such that:

$$(i') \qquad S_1 \perp\!\!\!\perp X | T$$

and

$$(ii') \qquad S_2 \perp\!\!\!\perp Y | S_1, X, T$$

**Proof:**
Starting with

$$A(x, y, T \cup S) = \sum_t \sum_{s_1} \sum_{s_2} P(y|x, t, s_1, s_2) P(s_1, s_2, t)$$

$(ii')$ permits us to remove $s_2$ from the first factor and write

$$\begin{aligned} A(x, y, T \cup S) &= \sum_t \sum_{s_1} \sum_{s_2} P(y|x, t, s_1) P(s_1, s_2, t) \\ &= \sum_t \sum_{s_1} P(y|x, t, s_1) P(s_1, t) \end{aligned}$$

while $(i')$ permits us to reach the same expression from $A(x, y, T)$:

$$\begin{aligned} A(x, y, T) &= \sum_t \sum_{s_1} P(y|x, t, s_1) P(s_1|x, t) P(t) \\ &= \sum_t \sum_{s_1} P(y|x, t, s_1) P(s_1, t) \end{aligned}$$

which proves the theorem.

Theorem 3 generalizes closely related theorems by Stone (1993) and Robins (1997), in which $T \cup S$ is assumed to be admissible (see also Greenland et al. (1999)). The importance of this generalization was demonstrated by several examples in Section 3. Theorem 3 on the other hand invokes only the distribution $P(x, y, z, t)$ and makes no reference to $P(y|do(x))$ or to admissibility.

Theorem 3 can also be proven by double application of Theorem 1; first showing the $c$-equivalence of $T$ and $\{T \cup S_1\}$ using (i) (with (ii) satisfied by subsumption), then showing the $c$-equivalence of $\{T \cup S_1\}$ and $\{T \cup S_1 \cup S_2\}$ using (ii) (with (i) satisfied by subsumption).

The advantage of Theorem 3 over Theorem 1 is that it allows certain cases of c-equivalence to be verified in a single step. In Figure 1, for example, both $(i')$ and $(i'')$ are satisfied for $T = \{V_1, W_2\}$, $S_1 = \{V_2\}$, and $S_2 = \{W_1\}$), Therefore, $T = \{V_1, W_2\}$ is c-equivalent to $\{T \cup S\} = \{V_1, V_2, W_1, W_2\}$

The weakness of Theorem 3 is that it is applicable to set-subset relations only. A natural attempt to generalize the theorem would be to posit the condition that $T$ and $Z$ each be c-equivalent to $T \cup Z$, and use Theorem 3 to establish the required set-subset equivalence. While perfectly valid, this condition is still not complete; there are cases where $T$ and $Z$ are c-equivalent, yet none is c-equivalent to their union. For example, consider the path

$$X \to T \leftarrow L \to Z \leftarrow Y$$

Each of $T$ and $Z$ leaves the path between $X$ and $Y$ blocked, which renders them c-equivalent, yet $\{T \cup Z\}$ unblocks that path. Hence, $T \approx Z$ and $T \not\approx T \cup Z$. This implies that sets $T$ and $Z$ would fail the proposed test, even though they are c-equivalent.

The remedy can be obtained by re-invoking the notion of Markov boundary (Definition 4) and Lemma 2.

**Theorem 4.** *Let $T$ and $Z$ be two sets of covariates, containing no descendant of $X$ and let $T_m$ and $Z_m$ be their Markov boundaries. A necessary and sufficient condition for the c-equivalence of $T$ and $Z$ is that each of $T_m$ and $Z_m$ be c-equivalent to $T_m \cup Z_m$ according to the set-subset criterion of Theorem 3.*

**Proof**

1. Proof of sufficiency:
   If $T_m$ and $Z_m$ are each c-equivalent to $T_m \cup Z_m$ then, obviously, they are c-equivalent themselves and, since each is c-equivalent to its parent set (by Lemma 2) $T$ and $Z$ are c-equivalent as well.

2. Proof of necessity:
   We need to show that if either $T_m$ or $Z_m$ is not c-equivalent to their union (by the test of Theorem 3), then they are not c-equivalent to each other. We will show that using "G-admissibility" as an auxiliary tool. We will show that failure of $Z_m \approx T_m \cup Z_m$ implies non-admissibility, and this, by the necessary part of Theorem 2, negates the possibility of c-equivalence between $Z$ and $T$. The proof relies on the monotonicity of d-separation over minimal subsets (Appendix B), which states that, for any graph $G$, and any two subsets of nodes $T$ and $Z$, we have:

$$(X \perp\!\!\!\perp Y | Z_m)_G \& (X \perp\!\!\!\perp Y | T_m)_G \Rightarrow (X \perp\!\!\!\perp Y | Z_m \cup T_m)_G$$

Applying this to the subgraph consisting of all back-door paths from $X$ to $Y$, we conclude that $G$-admissibility is preserved under union of minimal sets. Therefore, the admissibility of $Z_m$ and $T_m$ (hence of $Z$ and $T$) entails admissibility of $Z_m \cup T_m$. Applying Theorem 2, this implies the necessity part of Theorem 3.

Theorem 4 reveals the statistical implications of the $G$-admissibility requirement in Theorem 2. $G$-admissibility ensures the two c-equivalence conditions:

$$T_m \approx \{T_m \cup Z_m\} \qquad (7)$$

$$Z_m \approx \{T_m \cup Z_m\} \qquad (8)$$

In other words, given any DAG $G$ compatible with the conditional independencies of $P(x, y, t, z)$, whenever $Z$ and $T$ are $G$-admissible in $G$, the two statistical conditions of Theorem 3 should hold in the distribution, when applied to the set-subset relations in (7) and (8). Enforcing conditions $(i')$ and $(ii')$ with the proper choice of $S_1$ and $S_2$, yields

$$\{T_m \cup Z_m\}_m \perp\!\!\!\perp Y | X, Z_m \qquad (9)$$

$$\{T_m \cup Z_m\}_m \perp\!\!\!\perp Y | X, T_m \qquad (10)$$

which constitute the statistical implications of admissibility. These implications should be confirmed in any graph $G'$ that is Markov equivalent to $G$, regardless of whether $T$ and $S$ are $G$-admissible in $G'$.

We illustrate these implications using Fig. 2. Taking $T = \{W_2, V_2\}$ and $Z = \{V_1, W_3\}$, we have:

$$\begin{aligned} T_m &= \{W_2, V_2\},\ Z_m = \{V_1\}, \\ \{T_m \cup Z_m\}_m &= \{V_1, V_2, W_2\}_m = \{V_1, W_2\} \end{aligned}$$

We find that the tests of (9) and (10) are satisfied because

$$\{V_1, W_2\} \perp\!\!\!\perp Y | X, V_1 \quad \text{and} \quad \{V_1, W_2\} \perp\!\!\!\perp Y | X, W_2, V_2$$

Thus, implying $Z \approx T$. That test would fail had we taken $T = \{W_2\}$ and $Z = \{W_3\}$, because then we would have:

$$\begin{aligned} T_m &= \{W_2\},\ Z_m = \{\emptyset\}. \\ \{T_m \cup Z_m\}_m &= W_2 \end{aligned}$$

and the requirement

$$\{T_m \cup Z_m\}_m \perp\!\!\!\perp Y | X, Z_m$$

would not be satisfied because

$$W_2 \not\perp\!\!\!\perp Y | X$$

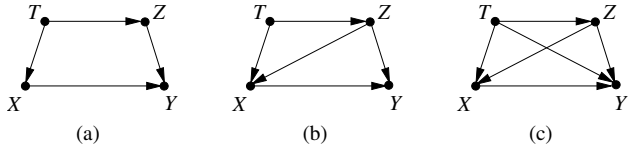

Figure 3: Failing the $T \approx \{T \cup Z\}$ test should reject Model (a) in favor of (b) or (c). Failing $Z \approx \{T \cup Z\}$ should reject Models (a) and (b) in favor of (c).

## 6 EMPIRICAL RAMIFICATIONS OF $c$-EQUIVALENCE TESTS

Having explicated the statistical implications of admissibility vis a vis $c$-equivalence, we may ask the inverse question: What can $c$-equivalence tests tell us about admissibility? It is well known that no statistical test can ever confirm or refute the admissibility of a given set $Z$ (Pearl, 2000, Chapter 6; Pearl, 1998). The discussion of Section 5 shows however that the admissibility of two sets, $T$ and $Z$, does have testable implications. In particular, if they fail the $c$-equivalence test, they cannot both be admissible. This might sound obvious, given that admissibility entails zero bias for each of $T$ and $Z$ (Eq. (5)). Still, Eq. (8) implies that it is enough for $Z_m$ (or $T_m$) to fail the $c$-equivalence test vis a vis $\{Z_m \cup T_m\}$ for us to conclude that, in addition to having different Markov boundaries, $Z$ and $T$ cannot both be admissible.

This finding can be useful when measurements need be chosen (for adjustment) with only partial knowledge of the causal graph underlying the problem. Assume that two candidate graphs recommend two different measurements for confounding control, one graph predicts the admissibility of $T$ and $Z$, and the second does not. Failure of the $c$-equivalence test

$$T_m \approx \{T_m \cup Z_m\} \approx Z_m$$

can then be used to rule out the former.

Figure 3 illustrates this possibility. Model 3(a) deems measurements $T$ and $Z$ as equally effective for bias removal, while models 3(b) and 3(c) deem $T$ to be insufficient for adjustment. Submitting the data to the $c$-equivalence tests of Eq. (7) and Eq. (8) may reveal which of the three models should be ruled out. If both tests fail, we must rule out Models 3(a) and 3(b), while if only Eq. (8) fails, we can rule out only Model 2(a) (Eq. (7) may still be satisfied in Model 3(c) by incidental cancellation).

Of course, the same model exclusion can be deduced from conditional independence tests. For example, Models 3(a) and 3(b) both predict $T \perp\!\!\!\perp Y | X, Z$ which, if violated in the data, would leave Model 3(c) as our choice and behoove us to adjust for both $T$ and $Z$. However, when the dimensionality of the conditioning sets increases, conditional independence tests are both unreliable and computationally expensive. Although both $c$-equivalent and conditional-independence tests can reap the benefits of propensity scores methods (see Appendix C) which reduce the dimensionality of the conditioning to a single scalar, it is not clear where the benefit can best be realized, since the cardinalities of the sets involved in these two types of tests may be substantially different.

This raises the interesting question of whether the discrimination power of $c$-equivalence equals that of conditional independence tests. We know from Theorem 4 that all $c$-equivalence conditions can be derived from conditional independence relations. The converse, however, is an open question if we allow $(X, Y)$ to vary over all variable pairs.

## 7 CONCLUSIONS

Theorem 2 provides a simple graphical test for deciding whether one set of covariates has the same bias-reducing potential as another. The test requires either that both sets satisfy the back-door criterion or that their Markov boundaries be identical. Both conditions can be tested by fast, polynomial time algorithms, and could be used to guide researchers in deciding what measurement sets are worth performing. We have further shown that the conditions above are valid in causal as well as associational graphs; the latter can be inferred from nonexperimental data. Finally, we postulate that $c$-equivalence tests could serve as valuable tools in a systematic search for graph structures that are compatible with the data.

### Acknowledgments

This research was supported in parts by grants from NIH #1R01 LM009961-01, NSF #IIS-0914211, and ONR #N000-14-09-1-0665.

## APPENDIX A

In this Appendix we prove the uniqueness of the Markov boundary $S_m$, as defined in Eq. (6), and provide a simple algorithm for its construction. Since $d$-separation is a graphoid (Pearl, 1988, p. 128), we will use the five graphoid axioms to prove uniqueness.

For a given $X$ and $S$ in a DAG $G$, assume $X$ has two different $S_m$'s, $U_1$ and $U_2$. Let $S_1$ be the intersection of $U_1$ and $U_2$; and let

$$U_1 = S_1 + S_2; U_2 = S_1 + S_3; S_4 = S - (U_1 + U_2).$$

Since $U_1$ and $U2$ are both Markov boundaries we have

$$(X \perp\!\!\!\perp S_2 S_4 | S_1 S_3)_G \text{ and } (X \perp\!\!\!\perp S_3 S_4 | S_1 S_2)_G \quad \text{(A-1)}$$

By Decomposition, we have

$$(X \perp\!\!\!\perp S_2 | S_1 S_3)_G \text{ and } (X \perp\!\!\!\perp S_3 | S_1 S_2)_G \quad \text{(A-2)}$$

By Intersection we get

$$(X \perp\!\!\!\perp S_2 S_3 | S_1)_G \quad \text{(A-3)}$$

From (A-1) and Weak Union we obtain

$$(X \perp\!\!\!\perp S_4 | S_2 S_3 S_1)_G \quad \text{(A-4)}$$

Now, applying Contraction to (A-3) and (A-4) we have

$$(X \perp\!\!\!\perp S_2 S_3 S_4 | S_1)_G \quad \text{(A-5)}$$

It follows that $S1$, which is smaller than $U1$ and $U2$ also satisfies Eq. (6) contradicting the minimality of $U1$ and $U2$

This completes the proof. (The case where $S_1$ is empty is included.)

It follows from the proof above that the Markov boundary $S_m$ is the intersection of all subsets of $S$ that $d$-separate $S$ from $X$. This implies that no member of $S_m$ can be $d$-separated from $X$ by any such subset. Therefore, the procedure of sequentially removing from $S$ any node that is $d$-separated from $X$ by all remaining nodes is guaranteed never to remove a member of $S_m$, hence it must terminate with the unique $S_m$.

(Note: The contraction axiom guarantees that removing any such node keeps all previously removed nodes $d$-separated from $X$ by the remaining nodes.)

## APPENDIX B

We prove that, for any graph $G$, and any two subsets of nodes $T$ and $Z$, we have:

$$(X \perp\!\!\!\perp Y | Z_m)_G \quad \& \quad (X \perp\!\!\!\perp Y | T_m)_G \Rightarrow (X \perp\!\!\!\perp Y | Z_m \cup T_m)_G$$

Where $Z_m$ and $T_m$ are any minimal subsets of $Z$ and $T$, that satisfy $(X \perp\!\!\!\perp Y | Z_m)_G$ and $(X \perp\!\!\!\perp Y | T_m)$ respectively.

The following notation will be used in the proof: A TRAIL will be a sequence of nodes $v_1, \ldots, v_k$ such that $v_i$ is connected by an arc to $v_{i+1}$. A collider $Z$ is EMBEDDED in a trail if two of his parents belong to the trail. A PATH is a trail that has no embedded collider. We will use the "moralized graph" test of Lauritzen et al. (1990) to test for $d$-separation ("$L$-test," for short).

**Theorem 5.** *Given a DAG and two vertices $x$ and $y$ in the DAG and a set $\{Z_1, \ldots, Z_k\}$ of minimal separators between $x$ and $y$. The union of the separators in the set, denoted by $Z!$, is a separator.*

**Proof.**
We mention first two observations:

(a) Given a minimal separator $Z$ between $x$ and $y$. If $Z$ contains a collider $w$ then there must be a path between $x$ and $y$ which is intercepted by $w$, implying that $w$ is an ancestor of either $x$ or $y$ or both. This follows from the minimality of $Z$. If the condition does not hold then $w$ is not required in $Z$.

(b) It follows from (a) above that $w$ as defined in (a) and its ancestors must belong to the ancestral subgraph of $x$ and $y$.

Let us apply the $L$-test to the triplet $(x, y | Z_1)$. As $Z_1$ is a separator, the $L$-test must show this. In the first stage of the $L$-test, the ancestral graph of the above triplet is constructed. By observation (b) it must include all the colliders that are included in any $Z_i$. In the next stage of the $L$-test, the parents of all colliders in the ancestral graph are moralized and the directions removed. The result will be an undirected graph including all the colliders in the separators $Z_i$ and their moralized parents and their ancestors. In this resulting graph, $Z_1$ still separates between $x$ and $y$. Therefore adding to $Z_1$ all the colliders in $Z_i$, $i = 1$ to $k$, will result in a larger separator. Adding the noncolliders from all the $Z_i$ to $Z_1$ will still keep the separator property of the enlarged set of vertices (trivial). It follows that $Z!$ is a separator. End of proof.

## Appendix C

Let the propensity score $L(z)$ stand for $P(X = 1 | z)$. It is well known (Rosenbaum and Rubin, 1983) that, viewed as a random variable, $L(z)$ satisfies $X \perp\!\!\!\perp L(z) | Z$. This implies that $A(x, y, L(z)) = A(x, y, Z)$ and, therefore, testing for the $c$-equivalence of $Z$ and $T$ can be reduced to testing the $c$-equivalence of $L(z)$ and $L(t)$. The latter offers the advantage of dimensionality reduction, since $L(z)$ and $L(t)$ are scalars, between zero and one. (See Pearl (2009a, pp. 348–352)).

The same advantage can be utilized in testing conditional independence. To test whether $(X \perp\!\!\!\perp Y | Z)$ holds in a distribution $P$, it is necessary that $(X \perp\!\!\!\perp Y | L(z))$ holds in $P$. This follows from the Contraction axiom of conditional independence, together with the fact that

$Z$ subsumes $L$. Indeed, the latter implies

$$X \perp\!\!\!\perp Y | Z \Leftrightarrow X \perp\!\!\!\perp Y | L(z), Z$$

which together with $X \perp\!\!\!\perp L(z)|Z$ gives

$$X \perp\!\!\!\perp Z | L(z) \ \& \ X \perp\!\!\!\perp Y | L(z), Z \Rightarrow X \perp\!\!\!\perp Y | L(z)$$

The converse requires an assumption of faithfulness.